\documentclass{article}
\usepackage{ismir,amsmath,cite}
\usepackage{graphicx}
\usepackage{color}
\usepackage{enumitem}
\usepackage{booktabs}
\usepackage[hang,flushmargin]{footmisc}

\title{Audio to score matching by combining phonetic and duration information}

\oneauthor
{Rong Gong, Jordi Pons and Xavier Serra}
{Music Technology Group, Universitat Pompeu Fabra, Barcelona\\ {\tt rong.gong@upf.edu, jordi.pons@upf.edu, xavier.serra@upf.edu}}

\begin{document}

	\maketitle
	\begin{abstract}
		We approach the singing phrase audio to score matching problem by using phonetic and duration information -- with a focus on studying the jingju a cappella singing case. We argue that, due to the existence of a basic melodic contour for each mode in jingju music, only using melodic information (such as pitch contour) will result in an ambiguous matching. This leads us to propose a matching approach based on the use of phonetic and duration information. Phonetic information is extracted with an acoustic model shaped with our data, and duration information is considered with the Hidden Markov Models (HMMs) variants we investigate. We build a model for each lyric path in our scores and we achieve the matching by ranking the posterior probabilities of the decoded most likely state sequences.
% 		given the phonetic information extracted by the acoustic model.
		Three acoustic models are investigated: \textit{(i)} convolutional neural networks (CNNs), \textit{(ii)} deep neural networks (DNNs) and \textit{(iii)} Gaussian mixture models (GMMs). Also, two duration models are compared: \textit{(i)} hidden semi-Markov model (HSMM) and \textit{(ii)} post-processor duration model. Results show that CNNs perform better in our (small) audio dataset and also that HSMM outperforms the post-processor duration model.
	\end{abstract}
	\section{Introduction}\label{sec:introduction}
	
	% what is singing phrase matching, what is the procedure and what is the result from that?
	
	\begin{figure}[h]
		\centering
		\includegraphics[width=8.5cm]{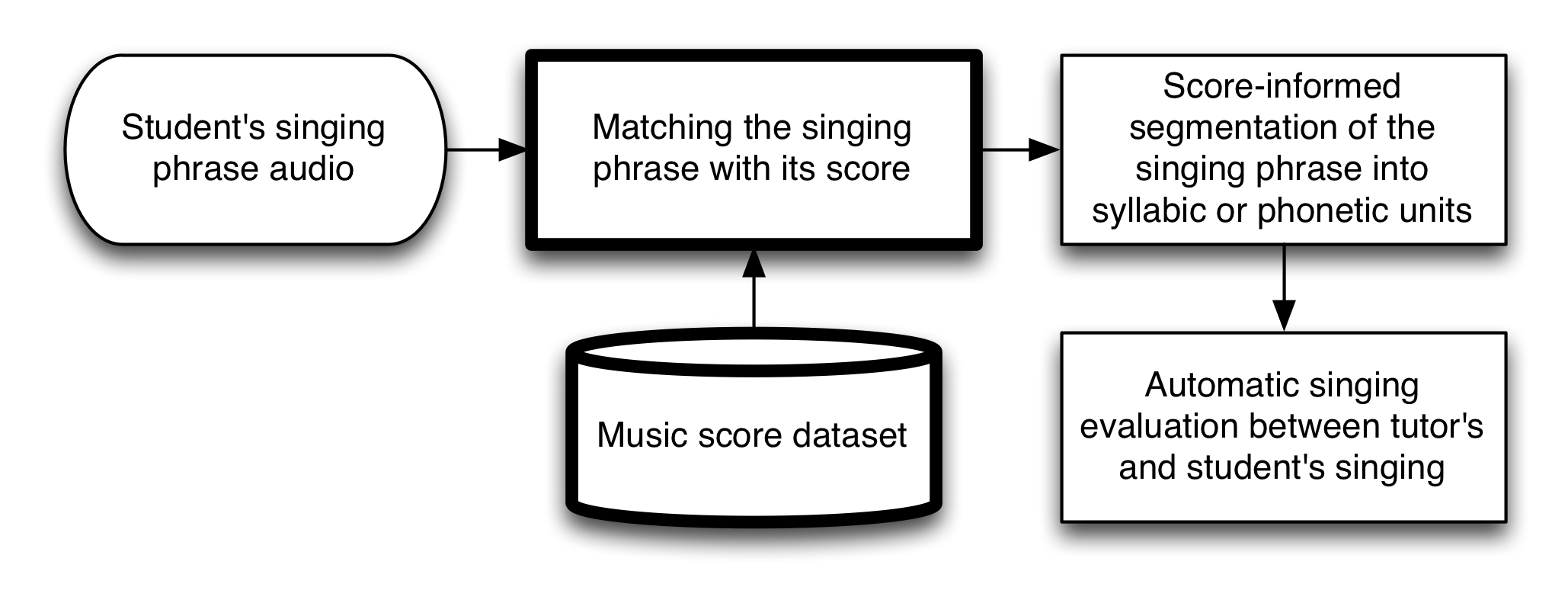}
		\caption{System design framework of the entire research project. The modules with bold border lines are addressed in this paper.}
		\label{fig:design_framework}
	\end{figure}
	
	% why we are interested in finding the matched score? what is this system used for? why not just using melodic similarity matching?
	The ultimate goal of our research project is to automatically evaluate the jingju a cappella singing of a student in the scenario of jingju singing education -- see Figure 1. Jingju, a traditional Chinese performing art also known as Peking or Beijing opera, is extremely demanding in the clear pronunciation and accurate intonation of each syllabic or phonetic singing unit. To this end, during the initial learning stages, students are required to completely imitate tutor's singing. Therefore, the automatic jingju singing evaluation system we envision is based on this training principle and measures the intonation and pronunciation similarities between the student's and the tutor's singings. Before measuring similarities, the singing phrase should be automatically segmented into syllabic or phonetic units in order to capture the temporal details. Jingju music scores, which contain the phonetic and duration information for each singing syllable, will be beneficial for this segmentation. In the application scenario, the score of a query audio could be selected from the database by the user itself. However, to avoid manual intervention and improve the user experience, we tackle the problem of automatically finding the corresponding music score for a given query audio (bold in \figref{fig:design_framework}). Note that achieving successful methods for audio to score matching might be beneficial for several music informatics research (MIR) tasks, such as: score-informed automatic syllable/phoneme segmentation \cite{gong2017score} or score-informed source separation \cite{Miron2016Score}.
	The objective of this research task is to find the corresponding score for a given singing audio query. We restrict this research to the ``matching" scope by pre-segmenting both the singing audios and the music scores into the phrase units.
	
\begin{figure}[h]
		\centering
		\includegraphics[width=8.5cm]{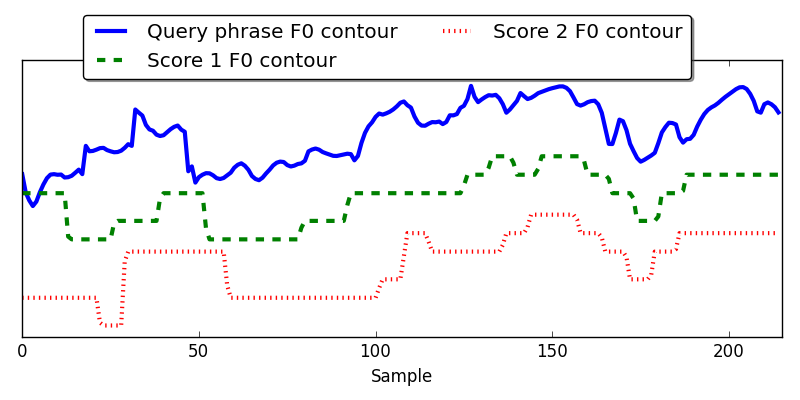}
		\caption{An example of different phrases having a similar \textit{Xipi} melodic contour in our score dataset. The lengths of these contours are normalized to a fixed sample size}
		\label{fig:melodic_contour_similar}
	\end{figure}

	\textit{Xipi} and \textit{erhuang} are the main modes in jingju music. Each has two basic melodic contours -- an opening phrase and a closing phrase. Each basic melodic contour is constructed upon characteristic pitch progressions for each mode \cite{wichmann1991listening}. Therefore, singing phrases from different arias sharing the same mode are likely to have a similar melodic contour. \figref{fig:melodic_contour_similar} shows an example of this fact. %that the F0 contours of two \textit{Xipi} scores can be found in our score dataset which are all similar to the F0 contour of the query phrase audio.
	% description of jingju shenqiang, show a figure which two candidate scores share the same melody. Motivate the exploration of phonetic and duration information.
	However, melodic information tends to be intuitively used for such matching tasks. For example in Query-by-Singing/Humming (QBSH) \cite{Molina2014TheIO}, melodic similarities can be obtained by comparing the distance between the F0 contour of the query audio and those synthesized from the candidate scores. Then, the best matched music score can be retrieved by selecting the most similar melody. But note that using this approach for jingju music would bring matching errors since the melodic contours of the same mode are similar in this sense. In this case, it is more appropriate to use another notion of similarity. We propose using the lyrics since the stories narrated in different jingju arias are distinctive and lyrics tend to change through different jingju arias -- even when they share the same mode. Therefore, phonetic information might be useful to identify a similar score given a query audio. 
	
	% related works: query by humming/singing by melodic similarity, keyword spotting, lyrics matching
	QBSH is the most related research task to our study, which retrieves a song by singing a portion of itself. Most of the studies use melody information as the only cue. The typical process of such systems was introduced by Molina \textit{et al.} \cite{Molina2014TheIO}: firstly, the F0 contour and/or a note-level transcription for a given singing query are extracted; and then, a set of candidate songs are retrieved from a large database using a melodic matcher module.
	The most successful QBSH system, which obtained the best results in MIREX 2016 contest, is based on the method of multiple similarity measurements fusion \cite{Wang2008QBSH}. This system proposed a melodic matcher which combines several similarities that are note-based and frame-based. The authors claim that the fusion mechanism improves the query performance because no similarity measurement is perfect. Therefore, information sources that are complementary to each other might be beneficial for this approach.
	Very few studies have explored the capability of the phonetic information for QBSH. Guo \textit{et al.} \cite{Guo2012QBSHLyric} and Wang \textit{et al.} \cite{Wang2015Lyric} both used a lyric recognizer based on Hidden Markov Models (HMMs). Their recognition networks\footnote{The topology of the HMM is defined by the recognition network.} were constructed with the phonetic information from the query candidates database. They used frame-based MFCCs to create the acoustic models with GMMs. Then, the Viterbi algorithm was executed over the recognition networks to either obtain the most likely phonetic state sequence (for Wang \textit{et al.} \cite{Wang2015Lyric}) or the posterior probability of each possible decoding path (for Guo \textit{et al.} \cite{Guo2012QBSHLyric}). The final score of a query candidate is either based on semantic similarity \cite{Wang2015Lyric} or based on the posterior probability of its corresponding lyrics \cite{Guo2012QBSHLyric}.
	
	Another research task related to our study is singing keyword spotting. The main goal of this task is to search for one or more keywords in a singing query. The system proposed by Kruspe \cite{Kruspe2015Keyword} searches for a specific singing keyword on the resulting phoneme observations. A keyword-filler HMMs is employed for this purpose. She used two phoneme duration models: the HSMM and the post-processor duration model.
	
	Finally, both phonetic and duration information extracted from the score have been extensively used in alignment-related tasks, such as audio-to-score alignment and audio-to-lyrics alignment. For example, Gong \textit{et al.} \cite{gong2015Align} construct a left-to-right HSMM using phonetic and duration information. 
	Or Dzhambazov \textit{et al.} \cite{dzhambazov2015modeling} use a similar approach for aligning polyphonic audio.	Analogously, the proposed approach explores the use of both phonetic and duration information (available in scores) to tackle the matching ambiguity problem existing in jingju music.
	
	The remainder of this paper is organized as follows: the used dataset is introduced in section \ref{sec:dataset}, section \ref{sec:approach} explains the modules of the proposed approach -- detailing how to incorporate phonetic and duration information. Experiments and results are reported in section \ref{sec:experiments}, and section \ref{sec:conclusion} concludes and points out future work.
	
	\section{Dataset}\label{sec:dataset}
	The jingju a cappella singing dataset is composed of two overlapping parts: \textit{(i)} audio and \textit{(ii)} score datasets. 
	
	The audio dataset \cite{black_automatic_2014} used for this study consists of two role-types singing: \textit{dan} (young woman) and \textit{laosheng} (old man). The \textit{dan} part of this dataset has 42 recordings sung by 7 singers and the \textit{laosheng} part has 23 recordings sung by 7 \textit{laosheng} singers. 
	%4 of the 15 \textit{dan} arias and 7 of the 23 \textit{laosheng} arias are chosen as the test set, which occupy around 25 percent of the annotated phoneme boundaries in each part of the dataset. The rest of them are used as the training set. 
	The boundary annotations of the audio dataset have been done in Praat format (textgrid) considering a hierarchy of three levels: phrase, syllable and phoneme -- using Chinese characters, pinyin notations and X-SAMPA notations, respectively. 32 phoneme classes are used in the phoneme-level annotation. Two Mandarin native speakers and a jingju musicologist have been devoted to this annotating work. Annotations and more detailed information can be found online\footnote{http://doi.org/10.5281/zenodo.344932}. Some statistics about the dataset are reported in \tabref{table:detailInfoDataset}. The average phrase, syllable and voiced phoneme length of \textit{dan} singing are ostensibly greater than those of \textit{laosheng} singing (bold numbers in \tabref{table:detailInfoDataset}), which might indicate that \textit{dan} singing tends to have more pitch variation and ornamentation -- as we could observe empirically by listening to the data.
	
	\begin{table}[!ht]
		\centering
		\caption{Detailed information of the jingju a cappella singing audio dataset: \textit{dan} (top), \textit{laosheng} (bottom).}
		\label{table:detailInfoDataset}
		\begin{tabular}{lccc}
			\toprule
			& Num. & Avg. len (s) & Std. len (s) \\
			\midrule
			Phrases & 325 & \textbf{16.42} & 14.11 \\
			Syllables & 2933 & \textbf{1.58} & 2.82 \\
			Voiced phonemes & 7198 & \textbf{0.61} & 0.97 \\
			Unvoiced phonemes & 2014 & 0.10 & 0.67\\
			\midrule
			Phrases & 247 & 9.47 & 8.14 \\
			Syllables & 2289 & 0.88 & 1.48 \\
			Voiced phonemes & 4948 & 0.39 & 0.78 \\
			Unvoiced phonemes & 1454 & 0.07 & 0.05\\
			\bottomrule
		\end{tabular}
	\end{table}
	
\begin{figure}[h]
		\centering
		\includegraphics[width=8.5cm]{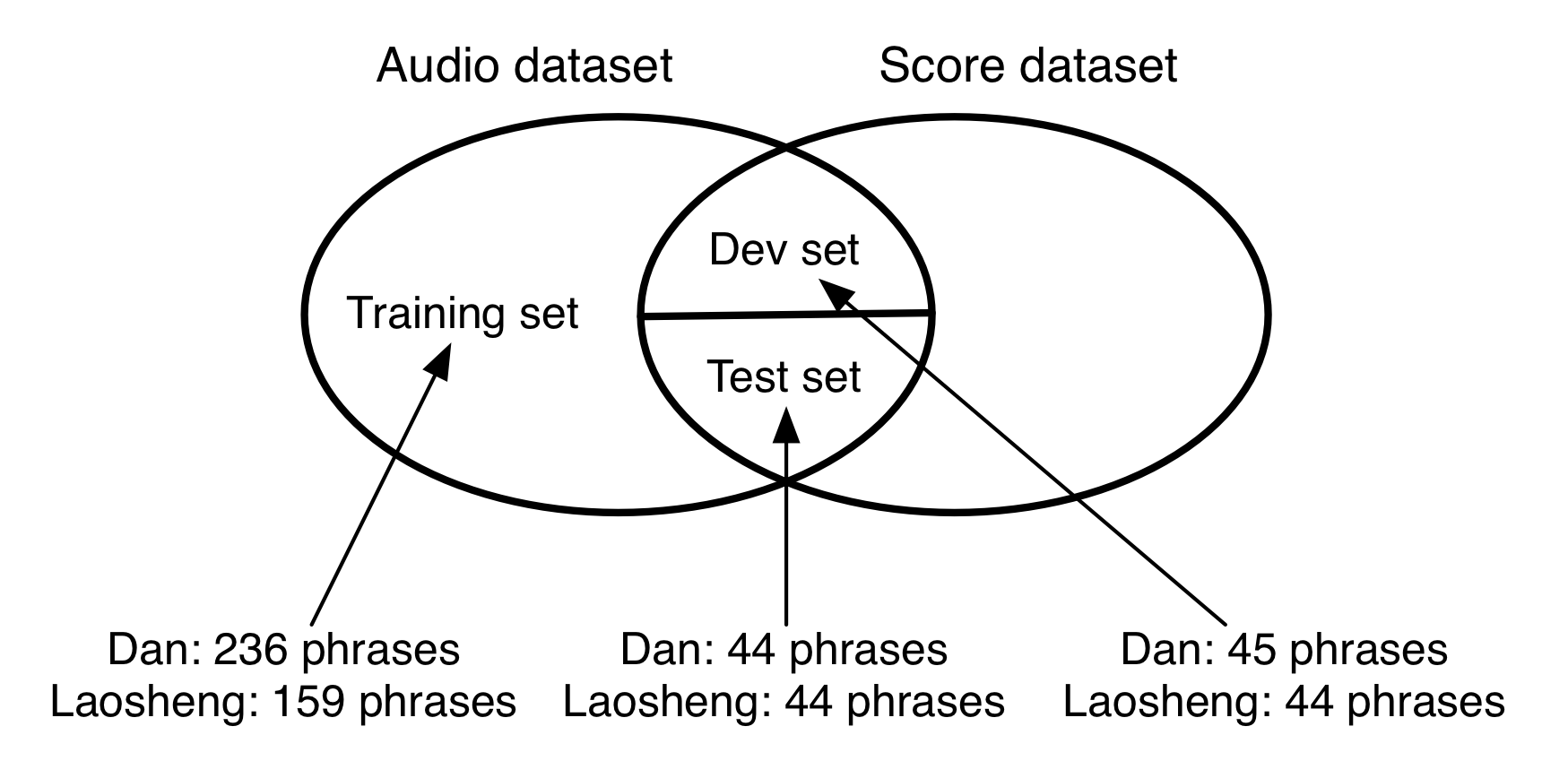}
		\caption{The intersection between the audio and the score datasets. The partition of the audio dataset.}
		\label{fig:dataset_intersection}
	\end{figure}

	The audio dataset, along with their boundary annotations, is split into three parts: training set, development (dev) set and test set. We define the training set to be the non-overlapping part with the score dataset, see \figref{fig:dataset_intersection}. The training set will be used for calculating the phonetic duration (duration information) and training acoustic models (phonetic information). 
	After taking the training set out, we define the development set to be the half of the remaining phrases in the audio dataset (randomly selected) -- it will be used for parameters optimization. The test set consists on the remaining phrases of the audio dataset -- it will be used for testing the acoustic models performance and the matching performance.	
	
	On the other hand, the score dataset contains 435 \textit{dan} phrases and 481 \textit{laosheng} phrases. The scores have been typed in stave notation (including lyrics in Chinese characters) using MuseScore from different printed sources in jianpu notation.  Since tempo is usually not clearly noted in the printed score, we do not include this information in the dataset. The relative syllabic durations are indicated by the note durations corresponding to the lyrics, which will be used to calculate the phonetic duration (duration information) and the matching network. The whole score dataset will be used as candidates for testing the matching performance and for parameter optimization.

\section{Approach}\label{sec:approach}
\begin{figure}[h]
    \centering
    \includegraphics[width=8.5cm]{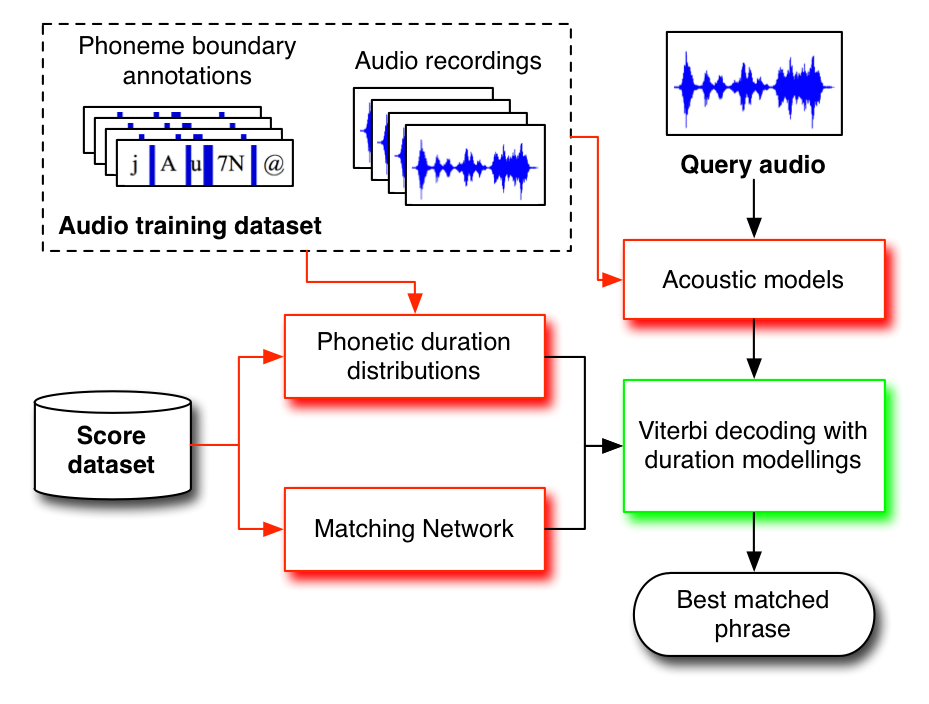}
    \caption{Diagram of the proposed approach.}
    \label{fig:main_diagram}
\end{figure}

The proposed approach aims to match the query audio to its score by using phonetic and duration information. During the training process (red boxes in \figref{fig:main_diagram}): the acoustic models of each phoneme are shaped by using the audio training set and its phonetic boundary annotations; the score dataset is used to construct a matching network; and phoneme duration distributions are estimated by using both audio training set and scores. During the matching process (green boxes in \figref{fig:main_diagram}): two duration models --HSMM and post-processor-- are explored for the Viterbi decoding step. Finally, the best matched phrase is found by ranking the decoded state sequence probabilities.

% \subsection{Role-type classification}\label{sec:role-type}
% Role-type classification of the query audio can be a effective pre-processing step to reduce the matching phrase candidate space. Since only \textit{dan} and \textit{laosheng} are considered as our query audio role-types, the opposite role-type phrase candidates can be removed from the matching space once the role-type of the query audio is identified. An example might be that if the query audio is \textit{dan} role-type, all the \textit{laosheng} phrase candidates will not be taken into account.

% We use XGBoost as the classifier because it requires not much parameter tuning and overfitting can be avoided by early-stopping. We extract a big amount of spectral, cepstral and tonal features by Essentia \cite{essentia2013} (such as spectral centroid, spectral spread, spectral energy, pitch confidence, pitch salience etc.) and compute commonly used statistical measures (e.g., mean, variance and standard deviation) from both the actual and the delta values as described in the paper \cite{Fuhrmann2010InstrRecog}. We normalize all features using $L2$ normalization and perform ANOVA F-test feature selection for the classifier.

\subsection{Acoustic models}\label{sec:acoustic_model}

Here presented acoustic models aim to represent the relationship between an audio signal and the 32 phoneme classes present in our dataset. The output of these models yield probability scores for each phoneme class.

The most popular way to approach acoustic modeling is by using GMMs and MFCCs features \cite{Guo2012QBSHLyric,Wang2015Lyric}. For that reason, we set as baseline a 40-component GMM with the following input vector: 13 MFCCs, their deltas and delta-deltas. Moreover, DNNs have been found very useful for acoustic modeling \cite{hinton2012deep,maas2017building}. Therefore, we propose an additional baseline: a DNN with 2 hidden layers followed by the 32-way softmax output layer -- the input is set to be a log-mel spectrogram. 

However, DNNs are very prone to over-fitting and the available dataset is relatively small. For that reason we propose using CNNs since these are more robust against over-fitting -- note that CNNs allow parameter sharing. Additionally, Pons \textit{et al.} \cite{pons2017timbre} have successfully used spectrograms-based CNNs for learning music timbre representations from small datasets. Given that timbre is an important feature for acoustic modeling, we propose using the same architecture: a single convolutional layer with filters of various sizes \cite{pons2017timbre,ponsdesigning}. The input is set to be a log-mel spectrogram. We use $128$ filters of sizes $50{\times}1$ and $70{\times}1$, $64$ filters of sizes $50{\times}5$ and $70{\times}5$, and $32$ filters of sizes $50{\times}10$ and $70{\times}10$  -- where the first and second numbers denote the frequential and temporal size of the filter, respectively. A max-pool layer of $2{\times}N'$ is followed by a 32-way softmax output layer with 30\% dropout -- where $N'$ denotes the \textit{temporal} dimension of the resulting feature map. $2{\times}N'$ max-pool layer was chosen to achieve time-invariant representations while keeping the frequency resolution. And \textit{same} padding is used to preserve the dimensions of the feature maps so that these are concatenable.
Filter shapes are designed so that filters can capture the relevant time-frequency contexts for learning timbre representations -- according to the design strategy proposed by Pons \textit{et al.} \cite{pons2017timbre}

Log-mel spectrograms are of size $80{\times}21$ -- the network takes a decision for a frame given its context: $\pm$10ms, 21 frames in total. Activation functions are ELUs \cite{clevert2015fast} for all deep learning models and these are optimized with stochastic gradient descent (batch size: 64), using ADAM \cite{kingma2014adam} and early stopping -- when validation loss (categorical cross-entropy) does not decrease after 10 epochs.

Spectrograms are computed from audio recordings sampled at 44.1 kHz. STFT is performed using a window length of 25ms (2048 samples with zero-padding) with a hop size of 10ms. The 80 log-mel bands energies are calculated on frequencies between 0Hz and 11000Hz and these are standardized to have zero mean and unit variance.

The acoustic models are trained separately for each role-type and their performance is reported in section \ref{sec:eval_am}.

\subsection{Matching network}\label{sec:matching_network}

\begin{figure}[h]
    \centering
    \includegraphics[width=8.5cm]{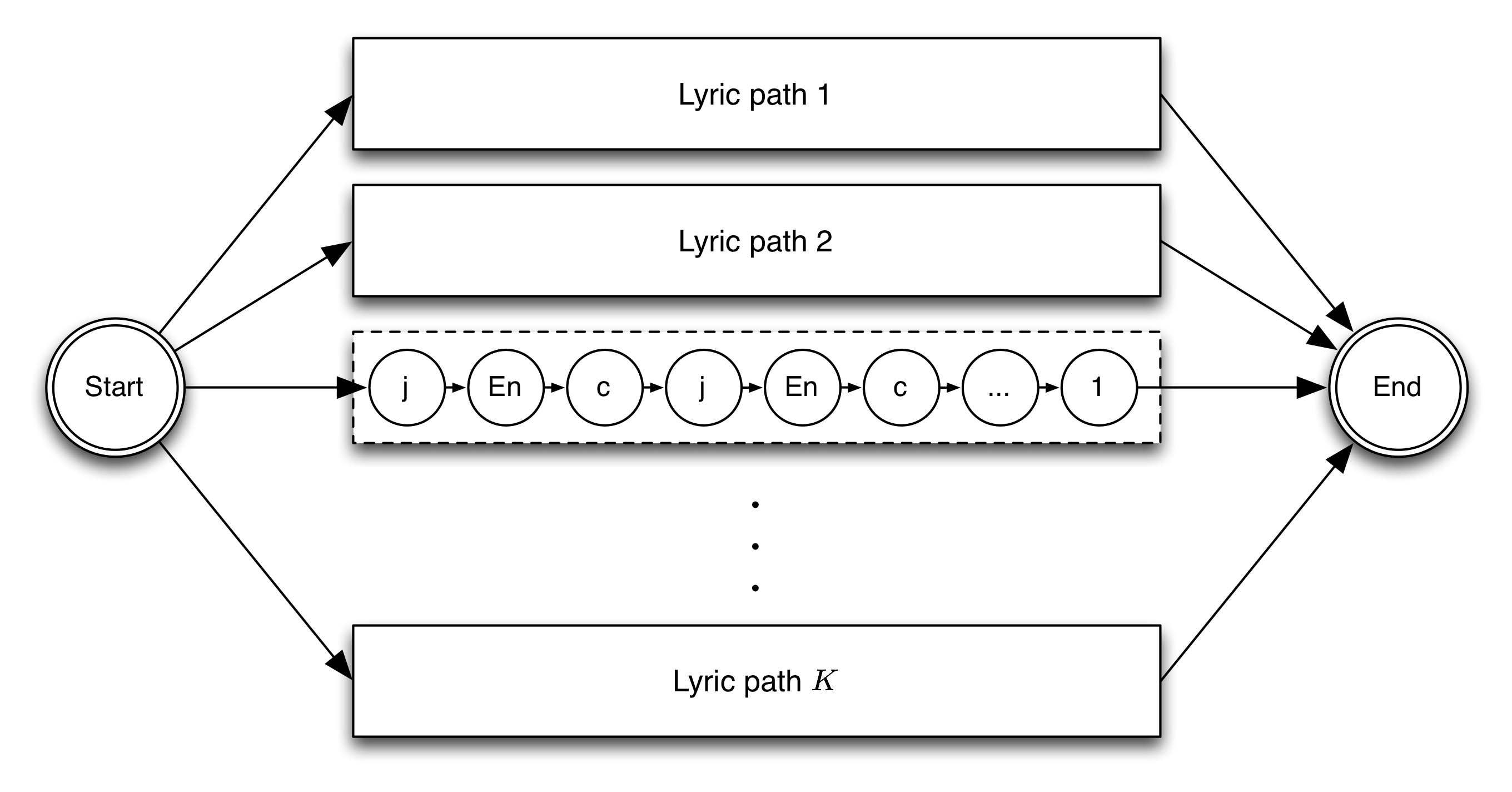}
    \caption{The structure of the $K$ paths isolated-phrase matching network. Path 3 shows an example of the left-to-right state chain structure of an individual lyric path.}
    \label{fig:matching_network}
\end{figure}

The matching network defines the topology of the hidden Markov model.
By using each candidate phrase in the score dataset as an isolated unit, isolated-phrase matching networks can be constructed. \figref{fig:matching_network} shows the structure of this matching network, which has $K=916$ lyric paths.

The matching network uses HMMs or HSMMs, depending on how the internal duration is modeled. Each path is a left-to-right state chain which represents the phoneme transcription of its lyrics. In order to construct the lyric path, pinyin lyrics are segmented into phonetic units and transcribed into X-SAMPA notations by using a predefined dictionary. For example, a path which has the lyrics \textit{yan jian de hong ri} in pinyin is a chain consisting of 12 states: j, En, c, j, En, c, 7, x, UN, N, r\textbackslash', 1 in X-SAMPA notation. When the decoding process has finished, each lyric path can get a posterior probability which will be used as the similarity measure between the query phrase and the candidate phrase.

%An isolated word recognizer performs better comparing with a continuous speech recognizer in the system, since the lyric of the input singing query is one of the K candidate lyrics. Due to the simplicity of the recognition network, the lyric recognition is fast and accurate.

\subsection{Phonetic duration distributions}\label{sec:dur_dist}
% phonetic information
% Phonetic information comes from two sources: the audio dataset boundary annotations and the score dataset. The process of extracting phoneme information is as follows: the lyrics (in Chinese characters) in the score is firstly converted to pinyin notations, then each pinyin is segmented into phonetic units and transcribed into X-SAMPA notations by using a predefined dictionary. 
%The phonetic X-SAMPA representation of each phrase candidate will be used to construct the matching network in section \ref{sec:matching_network}.

Phonetic duration information comes from two sources: the boundary annotations of audio training dataset and the score dataset. The phonetic duration is not directly indicated in the score. However, it is indispensable for modeling the phonetic duration distribution for each state in the matching network. The syllable, of which duration can be deduced by the corresponding note(s), is used to restrict the durations of the phonemes. 
\begin{figure}[h]
    \centering
    \includegraphics[width=8.5cm]{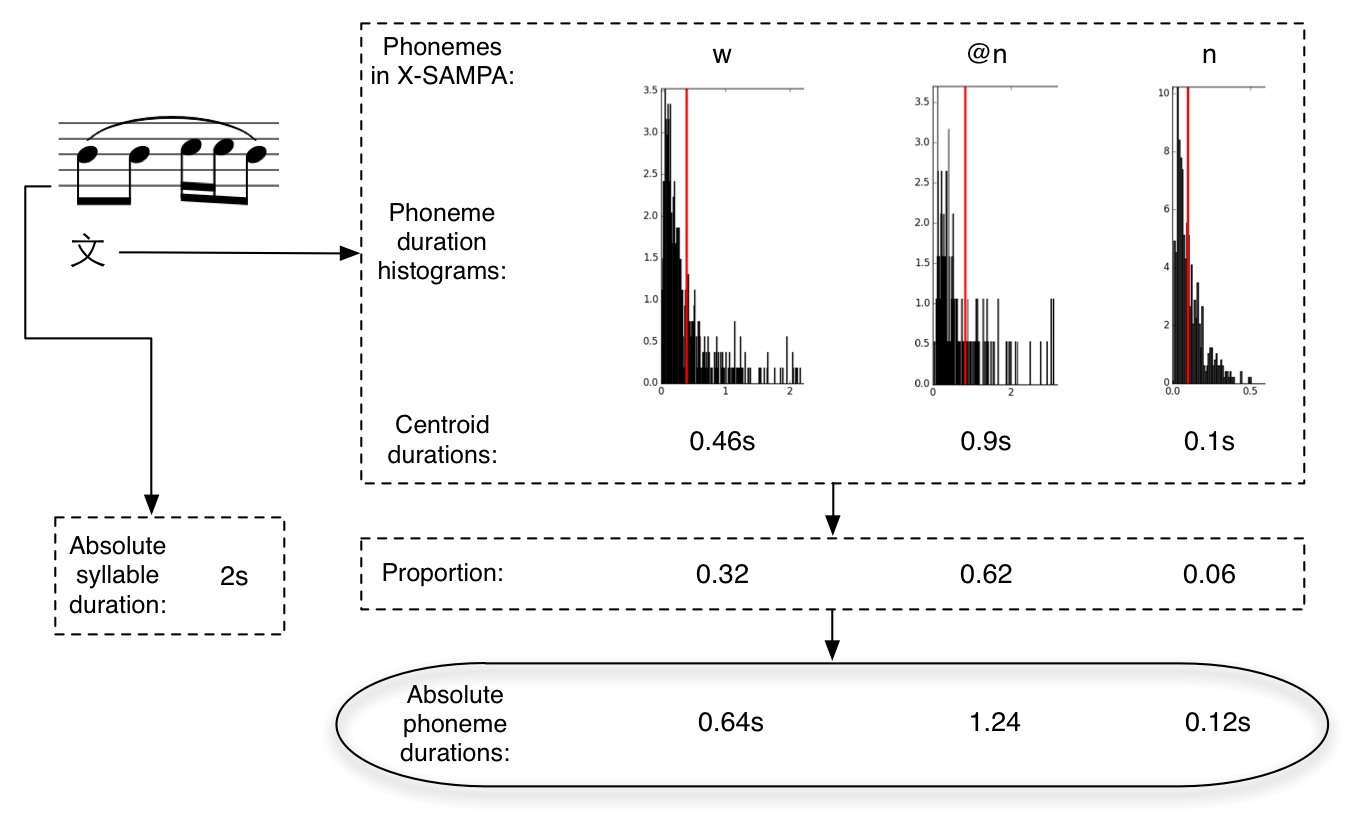}
    \caption{Flowchart example of estimating the phonetic durations of a syllable.}% The normalized absolute duration of this syllable in this example is 2s.}
    \label{fig:phoneme_duration}
\end{figure}

In the following, we propose a method for estimating the absolute phonetic duration given: \textit{(i)} the score, and \textit{(ii)} the phonemes duration histograms computed from the audio dataset annotations. First, we omit silence parts in the query audio (with a simple voice activity detection method \cite{sohn1999statistical}) and also in the score by removing the rest notes. Second, we compute the duration histogram and its duration centroid for each phoneme class -- by aggregating the phonetic durations indicated in the boundary annotations of the audio training dataset. 
Then, we segment each syllabic duration in the score dataset into phonetic durations according to the proportion of their duration centroids. Finally, as the scores do not contain tempo, we normalize the phonetic durations of each phrase such that their summation is equal to the duration of the query audio. See \figref{fig:phoneme_duration} for an equivalent graphic explanation.
In \figref{fig:phoneme_duration}, the centroid durations of these three phonemes are: 0.46s, 0.9s and 0.1s, summing: 1.46s -- alternatively, these can be expressed as a proportion of 1.46s: 0.32, 0.62 and 0.06. With these proportions and the absolute syllable duration (2s), we can compute the absolute phoneme durations: 0.32$\cdot$2s = 0.64s, 0.62$\cdot$2s=1.24s and 0.06$\cdot$2s=0.12s.

% See \figref{fig:phoneme_duration} for an alternative graphic explanation of the phonetic duration estimation process.

The phonetic duration distribution needs to be calculated for each state in the matching network in order to incorporate the \textit{a priori} phonetic duration information. 
%Burstein \cite{Burshtein1996duration} showed that Gamma distributions are the best at modeling English phoneme duration distributions. However, the Gamma distribution of each phoneme estimated from audio dataset can not be directly applied because the phonetic distribution should be also relevant to the syllabic duration exposed in the score. Thus, 
We model it by Gaussian distributions:
\begin{equation}
\label{eq:sylDurDist}
\mathcal{N} (x ; \mu_l, \sigma_l^2) = \frac{1}{\sqrt{2 \pi} \sigma_l} \exp \left(-\frac{(x-\mu_l)^2}{2\sigma_l^2} \right).
\end{equation}
where $\mu_l$ is the duration of the phoneme $l$ deduced by the above method and the standard deviation $\sigma_l$ is proportional to $\mu_l$: $\sigma_l=\gamma \mu_l$. The proportionality constant $\gamma$ will be optimized in section \ref{sec:optimization} for each role-type.

\eject %BREAK PAGE IN MULTICOLUMN!

\subsection{Duration modeling}\label{sec:duration_modeling}
Standard Markovian state do not impose explicitly duration distribution, instead, imposing an implicit state occupancy distribution which corresponds to a ``1-shifted" geometric distribution \cite{Guedon2005Hybrid}:
\begin{equation}
    d_j(u)=(1- \tilde{p}_{jj}) \tilde{p}_{jj}^{u-1}
\end{equation}
where $u$ denotes the occupancy or sojourn time in a Markovian state $j$ and $\tilde{p}_{jj}$ denotes the self-transition probability of the state $j$. Because of the implicity of the Markovian state occupancy, the phonetic duration distribution introduced in section \ref{sec:dur_dist} can not be imposed. Kruspe \cite{Kruspe2015Keyword} presents two duration modeling techniques for HMMs: Hidden semi-markov model (HSMM) and post-processor duration model.

\subsubsection{Hidden semi-markov model}
Gu\'{e}don \cite{Guedon2005Hybrid} defined a semi-Markov chain {$S_t$} with finite state space {$0,...,J-1$} by the following parameters:
\begin{itemize}[leftmargin=*]
    \item[-]initial probabilities $\pi_j=P(S_0=j)$ with $\sum_j \pi_j = 1$
    \item[-]transition probability of semi-Markovian state $j$: for each $k \neq j,p_{jk}=P(S_{t+1}=k|S_{t+1}\neq j,S_{t}=j)$ with $\sum_{k\neq j} p_{jk}=1$ and $p_{jj}=0$
\end{itemize}
An explicit occupancy distribution is attached to each semi-Markovian state:
\begin{equation}
\begin{aligned}
    d_j(u)=P(& S_{t+u+1}\neq j,S_{t+u-v}=j,v=0,...,u-2\\
    &|S_{t+1}=j,S_t\neq j), u=1,...,M_j
\end{aligned}
\end{equation}
where $M_j$ denotes the upper bound to the time spent in state $j$. $d_j(u)$ defines the conditional probability of leaving state $j$ at time $t+u+1$ and entering state $j$ at time $t+1$.

To apply HSMMs to the matching network, we first use the matching network as the HSMMs topology, thus the state occupancy distribution is set to its corresponding phonetic duration distribution. Then the probabilities of each left-to-right state transition are set to 1 because all self-transition probabilities in HSMMs are 0. The goal is to find the most likely sequence of hidden states for each lyric path and collect its posterior probability. The Viterbi algorithm meets this exact goal and its complete implementation is provided in \cite{guedon2007exploring}.

\subsubsection{Post-processor duration model}\label{sec:post_processor}
The post-processor duration model was first introduced by Juang \textit{et al.} \cite{Juang1985Isolated}. It was then experimentally proved in Kruspe's paper \cite{Kruspe2015Keyword} that this duration model works better than HSMMs for the keyword spotting task in English pop singing voice. The post-processor duration model uses the original HMMs Viterbi algorithm -- therefore, during the decoding process no explicit occupancy duration distribution is imposed.

The log posterior probability of the decoded most likely state sequence is augmented by the log duration probabilities:
\begin{equation}
    \log \hat{f} = \log f + \alpha \sum\limits_{j=1}^{N}\mathcal{N}(u_j; \mu_j,\sigma_j^2)
\end{equation}
where $f$ is the HMMs posterior probability, $\alpha$ is a weighting factor which will be optimized in section \ref{sec:optimization}, $j=1,...,N$ is the decoded state number in the most likely state sequence, and $\mathcal{N}(u_j; \mu_j,\sigma_j^2)$ is the occupancy probability of being in state $j$ for the occupancy $u_j$.

\section{Experiments and results}\label{sec:experiments}
\subsection{Performance metrics}
Two experiments\footnote{ Code:https://github.com/ronggong/jingjuSingingPhraseMatching/tree/v0.1.0} are performed: the first is to evaluate the performance of the acoustic models, and the second is to evaluate the proposed matching approaches. For the first task, we use one simple evaluation metric: the overall classification accuracy which is defined as the fraction of instances that are correctly classified. For the second task, our goal is to evaluate the ability of matching the ground-truth phrase in the score dataset to the query one, which is almost identical to the goal of a QBSH system: \textit{"finding the ground-truth song in a song database from a given singing/humming query"}. Therefore, we borrow the standard performance metrics used in QBSH task to evaluate our approaches: Top-M hit and Mean Reciprocal Rank (MRR) \cite{Guo2012QBSHLyric}. The Top-M hit rate is the proportion of queries for which $r_i\leq M$, where $r_i$ denotes the rank of the ground-truth score phrase. MRR is the average of the reciprocal ranks across all queries, $n$ is the number of queries, and ${rank}_i$ is the posterior probability rank of the ground-truth phrase corresponding to the i-th query. 

\begin{equation}
    MRR={\frac{1}{n}}\sum\limits_{i=1}^{n} {\frac{1}{rank_{i}}}
\end{equation}

\subsection{Acoustic models}\label{sec:eval_am}
CNN, DNN and GMM acoustic models yield probability scores for each phoneme class. In order to evaluate the classification accuracy, we choose the phoneme class with the maximum probability score as the prediction. \tabref{table:performance_am} reports the performance of CNN, DNN and GMM acoustic models evaluated on the test set. 
\begin{table}[!h]
	\centering
	\caption{Overall classification accuracies of CNN and baseline acoustical models for \textit{dan} and \textit{laosheng} datasets.}
	\label{table:performance_am}
	\begin{tabular}{lcccc}
		\toprule
		& \textit{dan}(\#parameters) & \textit{laosheng}(\#parameters) \\
		\midrule
		CNNs & 0.484(222k) & 0.432(222k)\\
		DNNs & 0.284(481k) & 0.282(430k) \\
		GMMs & 0.290(-) & 0.322(-) \\
		\bottomrule
	\end{tabular}
\end{table}

The relatively low classification accuracies for all three models show that modeling the phonetic characteristics for jingju singing voice is a challenging problem. Our best results are achieved with CNNs -- and GMMs perform better than DNNs. Interestingly, these results contrast with the literature where Hinton \textit{et al.} \cite{hinton2012deep} describe that DNNs acoustic models largely outperform GMMs for automatic speech recognition, and Maas \textit{et al.} \cite{maas2017building} showed that CNNs perform worse than DNNs for building speech acoustic models. First, we argue that in our case DNNs perform worse than GMMs and CNNs because a small amount of training data is available. DNNs require a lot of training data to achieve good performance and note that large amounts of training data are typically not available for most MIR tasks.  And second, note the CNNs used here are specifically designed to efficiently learn timbre representations \cite{pons2017timbre} while Maas \textit{et al.} \cite{maas2017building} used small squared filters, which proved successful
in computer vision tasks.
These results show that using CNN architectures designed for the task at hand is specially beneficial in small data scenarios.
A CNN model is used in the following experiments

\subsection{Parameters optimization}\label{sec:optimization}
The parameters which need to be optimized for \textit{dan} and \textit{laosheng} role-types are: the weighting factor $\alpha$ for the post-processor duration model, and the proportionality constant $\gamma$ for both models: HSMMs and post-processor duration model. 
%\textit{\color{red}The MRR metric can be reported by sweeping these parameters on the development set. Table \ref{tab:parameters_optimization} lists the search bounds and the optimal results.} 
Table \ref{tab:parameters_optimization} reports the optimal values we obtained by doing grid search on the development set -- MRR metric was maximized.

\begin{table}[ht!]
\centering
\caption{Parameters to be optimized, search bounds and resulting optimal values (\textit{dan} / \textit{laosheng}).}
\label{tab:parameters_optimization}
\begin{tabular}{llc}
\toprule
& \textbf{Search} & \textbf{Optimal}\\ 
\textbf{Parameters} & \textbf{bounds}  & \textbf{values}\\ \midrule
$\alpha$                        & $[0.25,2]$ with step 0.25     & 1.0 / 1.0    \\
$\gamma$ HSMMs                  & $[0.1,2]$ with step 0.1       & 0.1 / 0.1   \\
$\gamma$ post-processor         & $[0.1,2]$ with step 0.1       & 0.7 / 1.5   \\
\bottomrule
\end{tabular}
\end{table}

% \subsection{Role-type classification}
% The trained classifier is tested on the test audio dataset (\figref{fig:dataset_intersection}). The overall classification accuracy is 100\%, which means the role-type classification task is not challenging and the XGBoost classifier is reliable for the pre-processing step of identifying the role-type of the query audio and reducing the matching candidate space.

% \subsection{Baseline - Hidden markov models}
\subsection{Duration modeling}
\begin{figure}[h]
    \centering
    \includegraphics[width=7.92cm]{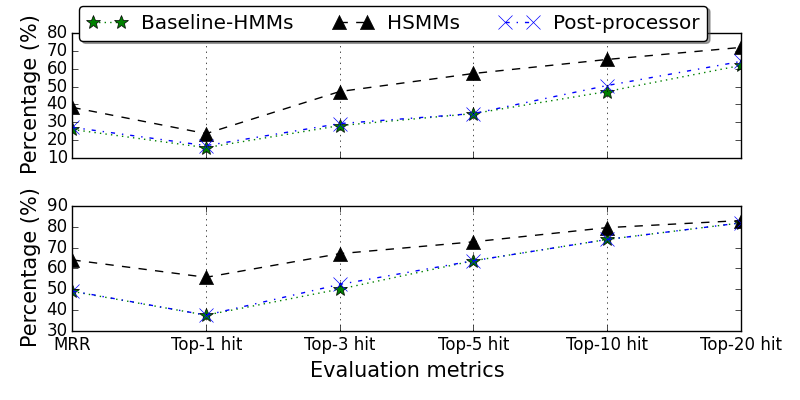}
    \caption{Phrase matching performance of HSMM and post-processor duration model with CNN acoustic model: \textit{dan} (top), \textit{laosheng} (bottom).}
    \label{fig:evaluation}
\end{figure}

To highlight the advantage of using duration modeling methods for audio to score matching, a standard HMM without explicitly imposing the occupancy distribution is used as a baseline. Results in \figref{fig:evaluation} show that its performance is inferior than the HSMM duration model.

% \subsection{Hidden semi-Markov models}
One can also observe in \figref{fig:evaluation} that HSMM performs the best, improving the baseline  MRR metric performance by 13.2\% for \textit{dan} role-type and 15.1\% for \textit{laosheng} role-type. This means that HSMMs explicit duration modeling can help achieve a better audio to score matching by using phonetic information.

% \subsection{Post-processor duration modeling}
The post-processor duration model does not significantly improve the baseline performance. This result contrasts with the literature, where the post-processor duration model worked better than HSMMs for singing voice keyword spotting \cite{Kruspe2015Keyword}.
This inconsistency might result from \textit{(i)} the length difference of the matching unit (singing-words vs. singing-phrases), and \textit{(ii)} the large standard deviation of the jingju singing phonemes length. 
First, in Kruspe's work \cite{Kruspe2015Keyword}, the matching unit is the singing keyword -- which usually contains less phonemes than a singing phrase (as in our case). And second, the vowel length standard deviation of the a cappella dataset used by Kruspe \cite{Kruspe2015Keyword} (around 0.3s) is much short than in our dataset (\textit{dan}: 0.97s, \textit{laosheng}: 0.78s) -- denoting less vowel duration variance than in our study case. Moreover, a significant deficiency of the post-processor duration model is that it does not provide the most likely state sequence by internally considering the durations, but it computes a \textit{new} weighted likelihood given the obtained sequence \cite{Kruspe2015Keyword}. If the most likely state sequence is decoded badly, it can't be restored by the post-processor duration model.

\section{Conclusions and future work}\label{sec:conclusion}
In this paper we presented an audio to score matching approach that uses phonetic and duration information.

We explored two duration models: HSMM and post-processor duration
model. HSMMs achieved better results than the post-processor duration
model -- probably due to \textit{(i)} the matching units length, \textit{(ii)} the large standard deviation of the considered phonemes,  and \textit{(iii)} because for the post-processor duration model it is hard to recover a decoding mistake. Moreover HSMMs achieved a better matching performance than the baseline-HMMs approach, which only took into account phonetic information, denoting the utility of using duration information.

We also compared CNN, DNN and GMM acoustic models, and CNNs have shown to be superior in our small singing voice audio dataset. The used CNN architecture was specifically designed to learn timbral representations efficiently \cite{pons2017timbre} -- this being the key factor for enabling CNNs (a deep learning method requiring large amounts of data) to perform so well in such a small dataset.

There are many possibilities to improve our approach. It has been shown in the speech research field that LSTM RNNs achieved the best acoustic modeling performance \cite{hasimRNNAM}. However, this method requires a large training dataset in order prevent from over-fitting. Another possibility to improve our acoustic model is to go deeper with the current single-layer CNN architecture, but this will also require more training data. We plan to collect more jingju a cappella singing recordings and perform data augmentation to leverage the capability of the acoustic models. Furthermore, in order to take advantage of the melodic information existing in both audio and score datasets, we also plan to investigate methods which can fuse melodic, phonetic and duration information.

\section{Acknowledgements}
We are grateful for the GPUs donated by NVidia. This work is partially supported by the Maria de Maeztu Programme (MDM-2015-0502) and by the European Research Council under the European Union's Seventh Framework Program, as part of the CompMusic project (ERC grant agreement 267583).

% For bibtex users:
\bibliography{ISMIRtemplate}

% For non bibtex users:
%\begin{thebibliography}{citations}
%
%\bibitem {Author:00}
%E. Author.
%``The Title of the Conference Paper,''
%{\it Proceedings of the International Symposium
%on Music Information Retrieval}, pp.~000--111, 2000.
%
%\bibitem{Someone:10}
%A. Someone, B. Someone, and C. Someone.
%``The Title of the Journal Paper,''
%{\it Journal of New Music Research},
%Vol.~A, No.~B, pp.~111--222, 2010.
%
%\bibitem{Someone:04} X. Someone and Y. Someone. {\it Title of the Book},
%    Editorial Acme, Porto, 2012.
%
%\end{thebibliography}

\end{document}